\documentclass{appolb}
\usepackage{epsfig}
\begin{document}
\title{Discussion session on diffractive Higgs production%
\thanks{Presented at DIS2002}%
}
\author{A. De Roeck
\address{CERN 1211 Geneva 23 Switzerland}
\and
 C. Royon
\address{CEA, SPP, DAPNIA, CE-Saclay, F-91191
Gif-sur-Yvette Cedex, France and
SPP and Texas U at Arlington (USA) }
}
\maketitle
\begin{abstract}
This note summarizes the discussion session on diffractive Higgs
production at the DIS2002 workshop.
\end{abstract}

\section{Introduction}
One of the main goals of the Large Hadron Collider (LHC) under 
construction at CERN is the search for,
discovery and measurement of the Higgs boson,
the particle associated with the field that can provide
a mechanism for electroweak symmetry breaking in the Standard Model.
The mass of the Higgs boson is unknown but precise measurements of
electroweak processes hint towards a value below 196 GeV
(95\% C.L.).
Direct searches exclude a Higgs with a mass smaller than 114.1 GeV
(95\% C.L.).
If Supersymmetry will turn out to be 
 the mechanism that stabelizes  the Standard Model at high energies, then the
 theoretically preferred region for the (lightest) Higgs mass is below 
135 GeV.

Measuring a light Higgs at the LHC will not be an easy task~\cite{deroeck},
but rather 
a delicate trade-off between signal and background. 
E.g. inclusive Higgs
production with the Higgs decaying into its most favourable mode, 
$b\overline{b}$,
cannot
be used to discover the Higgs due to the too high background of 
$b\overline{b}$ production. It is therefore important to explore
more, in particular clean, processes which would allow to discover the 
Higgs boson.

Recently, renewed attention has been drawn to diffractive Higgs 
production~\cite{rostovtsev}, being first discussed --as well as
heavy quark production --
in~\cite{landshoff}.
Since then, several groups have studied the processes but there are
substantial differences in the approaches used and 
results obtained. At the DIS02
meeting the most recent approaches were confronted in 
a discussion session. For simplicity
we will distinguish here only two 
main categories: {\it exclusive}  and {\it inclusive}
production, see Fig.~1. 
For each of these  there are several different
 models discussed. 
Furthermore we will only discuss those diffractive  channels 
where the incident protons survive the interaction and can be detected
in e.g. roman pot detectors, as these constitute the most interesting
event classes. But all these processes do have contributions from 
channels where the protons dissociate, and which can be used in case
one is only interested in the presence of a gap and not in
measuring the scattered proton.

\section{Exclusive production}
\begin{figure}[htb]
\centerline{\psfig{file=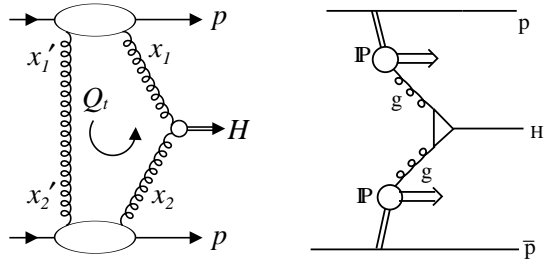,bbllx=0pt,bblly=500pt,bburx=210pt,bbury=710pt,width=4.5cm}
\psfig{file=higgs_remnant.eps,bbllx=0pt,bblly=-100pt,bburx=450pt,bburx=480pt,width=3.2cm}}
\caption{Diagrams for (left) exclusive and (right) inclusive diffractive 
Higgs production}
\label{fig1}
\end{figure}

In the case of exclusive production (Fig.~\ref{fig1}a), the final state is 
simple $pp\rightarrow p + H +p$. In such a configuration one can benefit
from strong spin  $J_Z =0$ selection
 rules which reduces
the LO order QCD background production such as 
$b\overline{b}$ production  by about 
two orders of magnitude.

The light Higgs production cross sections for exclusive production
 by the different calculations range from
approximately 100 fb \cite{landshoff,levin} 
to 3 fb \cite{martin}.
 Much of the difference between the results
comes from whether and how a so called
gap survival probability  is included in the calculation.
The Tevatron diffractive data imposes the need of a gap survival
probability of order 0.1 for most calculations of 
diffractive hard scattering processes.

The most detailed recent analysis of the exclusive channels is 
performed in~\cite{martin} and they 
find a cross section of the order of 3
fb. They also estimate --within their approach-- an uncertainty of a factor 
of two on this result~\cite{deroeck}.
In other words the exclusive Higgs production cross section could well
be rather small, but still detectable at the LHC.

\section{Inclusive processes}
Several groups have also studied the inclusive production cross section
(Fig.~\ref{fig1}b). Here we distinguish a so called 
factorizable, non-factorizable
and soft color interaction model.
A compilation of different recent calculations 
for the cross sections is given in 
Table~\ref{tab2}.

In the factorizable model~\cite{cox}, two pomerons are emitted with a structure
function and flux factor as measured in deep inelastic data at HERA.
The cross sections for both diffractive di-jet and Higgs production are 
calculated for Tevatron and LHC energies.
Diffractive di-jets have been measured at CDF, so the prediction can 
be compared to the data. The authors find this prediction
 a factor 10 too large, and 
rescale the Higgs cross section with this gap survival probability
accordingly. The cross sections for LHC in their paper are not 
rescaled.
In~\cite{martin} similar cross sections are calculated, and results 
similar results obtained,
be it using actually  different diagrams as explained in~\cite{deroeck}

In the non-factorizable model~\cite{royon,boon2}, two pomerons are emitted 
with a structure
function as measured in deep inelastic data at HERA, but a soft
flux factor ($\epsilon = 0.08 $ instead of $\sim 0.2$) is used, meant
to absorb the factorization breaking seen between HERA and Tevatron
hard diffractive measurements.
Similarly the cross sections for 
both diffractive di-jet and Higgs production are 
calculated for Tevatron and LHC energies.
The di-jet cross section is  found to be a factor 3.8 too small
at the Tevatron when calculated with conventional parameters, and 
the Higgs cross section is rescaled accordingly, also for the LHC
results. The resulting numbers are similar to~\cite{cox}, but 
for the  latter no gap suppression factor has been applied for the 
LHC predictions. If this is applied  the result 
in~\cite{cox} would be a factor
10-20 smaller than these results.

In~\cite{ingelman} the authors use the soft colour interaction (and also
the general area law) model to predict Higgs production cross sections.
This model can describe a variety of
  diffractive data at the Tevatron and HERA.
It predicts a small cross section for diffractive Higgs production at the 
Tevatron.

\begin{table}
\begin{center}
\begin{tabular}{|c|c|c|}
\hline
$\sigma_H $(fb) & Normalization & Ref. \\
\hline
320 & x 3.8 & \cite{boon2} \\
260-390 & no rescaling  & \cite{cox}\\
0.19-0.16 &  & \cite{ingelman}\\
\hline
\end{tabular}
\caption{Cross sections for inclusive Higgs production}\label{tab2}
\end{center}
\end{table}

\begin{figure}[htb]
\centerline{\psfig{file=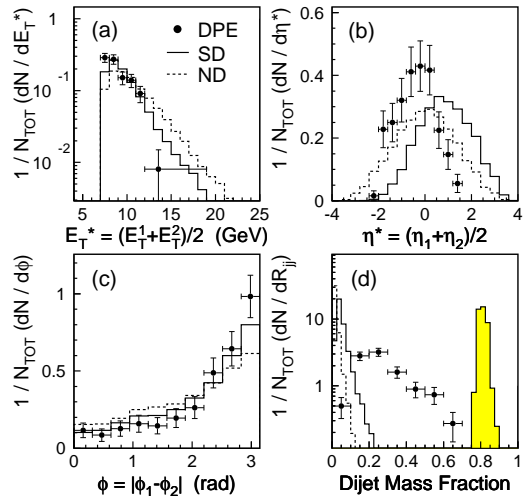,bbllx=0pt,bblly=30pt,bburx=560pt,bbury=540pt,width=7cm}
}
\caption{Di-jet  quantities DPE events as measured by CDF.
The bottom-right plot shows the dijet mass fraction}
\label{fig2}
\end{figure}

\section{Discussion}
The following discussion developed at the meeting.
\begin{itemize}

\item Comparison between the different calculations, 
especially the exclusive channels:
what is the origin of the differences in the models?
How can we control the all important gap survival probability
experimentally?

For the pomeron processes, the difference in predictions is a factor 
10 for those models that 'use' emition of 
pomerons. It appears that this factor can be 
mostly explained by the different value $\epsilon$ used for the 
flux~\cite{boonekamp}.

\item What are the uncertainties on the calculations?
By varying the parameters in the models for the cross section calculations
(input structure functions etc.) one could get a handle on the spread
within a given model.
One author of ~\cite{landshoff} reported that the uncertainty  can be easily
as large as a factor 10. Recently in~\cite{deroeck} the uncertainty 
was evaluated to be a factor two only for the calculations first
presented in~\cite{martin}

\item How can one test these models?
While now generally accepted that the diffractive Higgs 
production rates for the Tevatron are probably too small to be of use,
there are several processes which can be exploited
at this collider in the next years, e.g.
diffractive di-jet, di-photon, $\chi_c$ and possibly $\chi_b$ production.
Di-photon production should be measurable in run-II according to the 
predictions in~\cite{cox}.
Perhaps also vector meson production at HERA can play a role in
discriminating the models.

In particular diffractive di-jet production is very interesting.  
CDF~\cite{CDF} has already measured double pomeron
exchange (DPE) dijet production. Fig.~\ref{fig2}
shows --among other distributions-- 
the fraction of the energy of the dijets compared to the total 
energy in the central system. Clearly every inclusive model for Higgs 
production, when applied to predict Higgs and di-jet production rates should
describe the shape and normalization of this distribution.
Note however that the CDF data are not corrected for detector 
smearing, and thus to reproduce these signals one needs an event
generator and detector simulation which has the proper energy 
smearing.

CDF also sets a limit on the cross section for exclusive di-jet production
i.e. where all visible energy enters in the two jets.
They find that a most 5.1 events ( 3.7 nb) are compatible
with this  hypothesis. E.g. any model predicting a LARGER cross section 
than this can already be  excluded.

\item Concerning the debate of exclusive and inclusive production:
 What can one finally gain from the inclusive diffractive Higgs production
with respect to inclusive Higgs production?
In this case there will be no $J_Z$ selection rule to suppress the background
and one cannot use the relation
$M_{pp}=M_{H}$, since there are always remnants around. Some initial ideas
have been proposed in~\cite{boon2}, but need to be substantiated
with real hadronization and detector simulation.
In particular inclusive production studies should make 
a full background calculations to show that the signal will be visible
at the end.

There were concerns expressed whether
 exclusive events at such large scales really happen in nature.
  Will there not always be some soft gluons around which spoil the 
  exclusiveness? Di-photon production at the Tevatron would be  
  a good testing ground to confirm that these events are produced
  at high energies.

\end{itemize}

\section{Suggested homework}

\begin{itemize}
\item
The new Tevatron run-II data will be of pivotal importance to settle some of 
these questions. We suggest that the following data be collected.

Measure the DPE dijet spectra, preferably with 
double proton tagging to really constrain the $M_{dijet}$, and measure it
for different $E_T$ scales (such that one can test the $\epsilon$ value
of the flux).

 Try to measure the exclusive di-photon or a $\chi$ states. These 
have the advantage over the di-jets that it is easier to determine
their 'exclusiveness'. The cross sections are however much lower, so 
here the 
Run-II luminosity will be needed.

\item For the different models it would be useful to have the 
comparisons of the predictions with Fig.\ref{fig2}.
Predictions for higher jet $E_T$ cuts, such as 10 and 15 GeV would be 
useful for future comparisons with data and to demonstrate the 
cross section behaviour with the scale in the model.

\item For the different models it would be useful to have predictions for 
di-photon production rates, e.g. for photons with $E_T > 7 $ GeV, as
in~\cite{cox}.

\item A Monte Carlo generator for all these processes would be useful,
to compare with experimental data, eg. the di-jet mass fraction.
\end{itemize}

The goal is to have these model numbers available by the Low-$x$
meeting in September 2002 (Antwerpen/Belgium).

We thank all contributers to this session for their presentation and 
a lively discussion. Furthermore we thank R. Peschanski and V. Khoze 
for a critical reading of the paper.

\end{document}